\documentstyle[aps,epsfig,preprint]{revtex}
\draft
\begin{document}
\title{Electron-phonon scattering at the intersection of
two Landau levels\vspace{-15pt}}
\author{
V.~N.~Golovach and M.~E.~Portnoi \\ {\it School of Physics, 
University of Exeter, Stocker Road, Exeter, EX4 4QL, UK}\vspace{-20pt}}
\maketitle
\begin{abstract}\vspace{-30pt}\\
We predict a double-resonant feature in the magnetic field 
dependence of the phonon-mediated longitudinal conductivity $\sigma_{xx}$ 
of a two-subband quasi-two-dimensional electron system in a quantizing 
magnetic field. 
The two sharp peaks in $\sigma_{xx}$ appear when the energy separation 
between two 
Landau levels belonging to different size-quantization subbands 
is favorable for acoustic-phonon transitions. 
One-phonon and two-phonon mechanisms of electron conductivity are calculated
and mutually compared.
The phonon-mediated interaction between the intersecting Landau 
levels is considered and no avoided crossing is 
found at thermal equilibrium.
\end{abstract}
\section{INTRODUCTION}
The acoustic phonon scattering of electrons in two-dimensional (2D) systems 
in quantizing magnetic fields has been extensively studied for the 
last three decades. The peculiar density of states of the 2D electrons 
in a transversal magnetic field $B$ produces a suppresion of the 
electron-phonon stattering rate at strong magnetic 
fields~\cite{ref:1,ref:3}, 
at which the so called {\it inelasticity} parameter $\eta=s/\lambda\omega_c$ 
is less than unity, here $s$ is the sound velocity, $\lambda=\sqrt{\hbar/eB}$ 
is the magnetic length, and $\omega_c=eB/m$ is the cyclotron frequency. A 
detailed analysis of the acoustic phonon emission and absorption spectra in 
2D electron systems has been carried out in connection with developing a 
phonon absorption spectroscopy~\cite{ref:KB,ref:1a,ref:1b}, which is used 
to investigate the states of electronic matter in both integer and fractional 
quantum Hall effects. It was also noticed that, at a certain stage of 
supression 
of the one-phonon inter-Landau-level transition, the two-phonon transition 
dominates the electron relaxation rate~\cite{ref:Falko}. 
However, in a
two-subband system, the electron relaxation rate can still be due to 
the one-phonon transitions occurring between Landau levels of different
size-quantization subbands. An oscillatory behavior of the electron 
lifetime was found in a two-subband system~\cite{ref:Maksym}.

In this paper we focus on an intersection of two Landau levels belonging to
different size-quantization subbands of a two-dimensional electron gas in 
a transversal magnetic field $B$ (see Fig.~1).
The electron density is chosen so that there are enough electrons to fill
only one Landau level out of the two at the intersection point. 
Despite the general suppression due to small $\eta$, we find that,
at an energy separation between the two Landau levels of the order 
of the characteristic acoustic-phonon energy $T_{\lambda}=\hbar s/\lambda$, 
the electron-phonon scattering is significantly enhanced.
In what follows, we neglect the effects related to the Coulomb 
interaction between the electrons. These effects change significantly the 
results at low temperatures~\cite{Apalkov_Portnoi}. 
The virtue of the considered structure, from the point of view of 
electron-phonon scattering, is that 
an enhancement of the dissipative conductivity $\sigma_{xx}$ arises on the 
background of its strong overall suppression. This enhancement can be 
understood from the following physical consideration.

At the intersection of two Landau levels, say level $\ell=0$ of the upper
subband and level $\ell=1$ of the lower subband, the matrix element of the
electron-phonon interaction requires a phonon momentum $q\sim 1/\lambda$ for
the most efficient transition. Note that two Landau levels can be intersected
at a magnitude of $B$ corresponding to $\lambda\sim a$, where $a$ is the
characteristic width of the quantum well confining the two-dimensional 
electron gas. This results in $q_z\sim q_{\bot}\sim 1/\lambda$, where
$q_z$ and $q_{\bot}$ are the perpendicular and in-plane components of the
phonon momentum, respectively. 
On the other hand, the energy conservation law
requires that the phonon energy $\hbar\omega_{\bf q}$ matches the energy 
separation between the two Landau levels 
$\Omega=|{\cal E}_{\ell=0}-{\cal E}_{\ell=1}|$. 
For acoustic phonons, 
we have $\omega_{\bf q}=sq$, and if combined with the momentum requirement,
two {\em resonances} are obtained, one on each side of the crossing point,
corresponding to $\Omega\simeq \hbar s/\lambda\equiv T_\lambda$.
At each resonance the electron-phonon scattering is significantly enhanced,
owing to the one-phonon transitions.
 
It is worth noting that a simultaneous intersection of two Landau levels with 
the Fermi level leads to a missing quantum Hall effect plateau as expected
from a single-subband consideration.
Such intersections occur repeatedly with decreasing $B$, if 
the electron density per spin orientation $n_{2D}=n_c(1+(2p-1)/q)$, 
where $p$, $q$ are natural numbers, and $n_c=mE_{12}/2\pi\hbar^2$, with 
$E_{12}$ being the energy separation between the two 
size-quantization subbands. 
However, if the Coulomb interaction between the electrons is taken into 
account, a {\em new} energy gap emerges at the intersection 
point~\cite{Apalkov_Portnoi}.

An enhancement of the electron scattering in a two-subband system 
occurs also at zero magnetic field, when the second size-quantization
subband starts being filled with electrons. This effect can be viewed in terms
of a Lifshitz topological transition of the Fermi surface, which is well known
for the 3D case~\cite{ref:6}. The 2D case of this transition was 
considered in Ref.~\onlinecite{ref:7}, where the electron scattering occured 
on a random potential of impurities. At $B\neq 0$, one can hardly speak of a 
Fermi surface. However, we refer the peculiar behaviour 
of $\sigma_{xx}$ in our case to the same origin -- a strong variation in the 
density of states at the Fermi level. The case considered in this paper 
has the advantage of more pronunced anomalies.

An experimental realization of the above described situation can be acheived 
in GaAs/AlGaAs heterojunctions. Usually, the donor supply of the electrons
into the 2D layer is not sufficient to achieve filling of the second 
size-quantization subband. However, a 
significant increase in the concentration can be achieved by illuminating 
the sample with photons of energy close to the band gap of GaAs. The 
photoexcited electrons come either from the DX centers in the AlGaAs 
layer~\cite{ref:8} or from the valence band of the bulk GaAs producing a 
charge separation at the interface~\cite{ref:9}. Furthermore, the 2D electron 
concentration can be tuned within a large range using the method of continuous 
photoexcitation~\cite{ref:10}. Population of the second subband of size 
quantisation in a magnetic field has been observed in a number of experiments 
related to magneto-optical studies of the integer and fractional quantum 
Hall effects~\cite{ref:11}$^-$\cite{ref:13}.

In this paper we consider only scattering of electrons by phonons, 
although in a real 2D system the main contribution into $\sigma_{xx}$ at low 
temperatures comes from the scattering of electrons by the impurity potential. 
This was extensively studied in connection with the integer quantum Hall 
effect. However, since the electron-phonon coupling constant $\alpha\sim B$, 
 the phonon-induced $\sigma_{xx}$, at high enough magnetic fields, may be 
comparable with the impurity induced $\sigma_{xx}$.

%The two-phonon electron scattering mechanism within the lowest Landau level 
%was explicitly cosidered in Ref.~\onlinecite{ref:4}, however the paper 
%contains a mistake in the expression for the diffusion coefficient (Eq.(2) of 
%Ref.~\onlinecite{ref:4}), 
%which leads to wrong results at temperatures $T>\hbar s/\lambda$. In this 
%paper we intend 
%to give a full description of the electron dissipative conductivity due to 
%the two-phonon scattering mechanism. We take into account both 
%deformation-potential and piezoelectric types of electron-phonon interaction, 
%which was not done in Ref.~\onlinecite{ref:4}.             
     
\section{GENERAL RELATIONS}
We consider an electron gas in a quantum well, formed by a confining potential 
$V(z)$, in a strong magnetic field $B$ directed along the axis $z$. In the 
Landau gauge the vector potential ${\bf A}=(0,xB,0)$, and the electron can be 
characterised by the center of orbit coordinate $X$. The energy 
spectrum of the electron, ${\cal E}_{n\ell}=E_n+\hbar\omega_c(\ell+1/2)$, 
consists of the size quantization energy $E_n$ and the magnetic 
quantization energy, described by the Landau level number $\ell$. The acoustic 
phonons, with dispersion $\omega_{{\bf q}\,j}$, interact 
with the electrons via both deformation potential (DP) and piezoelectric (PE) 
mechanisms. We shall restrict ourselves to the isotropic Debye approximation 
in which the phonon modes fall into a branch of longitudinal (LA) phonons 
($j=1$) and two branches of transversal (TA) phonons ($j=2,3$). This seems to 
be a good approximation for phonons at thermal equilibrium, in contrast with 
the case of balistic phonon propagation, where the anysotropy gives rise to 
self-focusing effects~\cite{ref:Wolfe}. For GaAs, it follows that the DP 
mechanism couples electrons only with LA phonons. We assume that acoustic 
phonons are the only sourse of scattering for the electrons in the well, 
neglecting the 
effects of interface roughness and random potential of impurities, as well 
as the presence of optical phonon modes. This assumption is valid for fairly 
pure samples at temperatures below the optical phonon energy. We also suppose 
the acoustical phonons to be three dimentional, neglecting the effects of 
interface phonons. The electrons are considered spinless. The 
spin degeneracy can be easily accounted for at the final stage 
of calculation. Electron scattering by phonons is spin preserving, and the 
results for the conductivity will differ only by a factor of 2. 

The system of interacting electron and phonon gases is described by the 
following second quantized Hamiltonian
\begin{equation}
H=H_{0}+H_{int}\;\;,
\label{eq:2a}
\end{equation}
with
\begin{equation}
H_0=\sum_{n\, \ell\, X}{{\cal E}_{n\ell}C^{\dag}_{n\ell X}C_{n\ell X}}\;+
\sum_{{\bf q}\,j}{\hbar\omega_{{\bf q}\,j}(b^{\dag}_{{\bf q}\,j}
b_{{\bf q}\,j}+1/2)}\;\;,
\label{eq:2}
\end{equation}
and
\begin{equation} 
H_{int}=-\sum_{ n\ell n'\ell'\atop {\bf q}\,jX}
M^{{\bf q}\,j}_{X}
\!\!\,\left(\!\! 
\begin{array}{c}
n'  \ell' \vspace{-8 pt} \\
n \: \ell
\end{array}
\!\!\right)\,(b^{\dag}_{-{\bf q}\,j}+b_{{\bf q}\,j}) 
C^{\dag}_{n'\ell'X-\lambda^2q_y}C_{n\ell X}\;\;.
\label{eq:3}
\end{equation} 
Here $C^{\dag}_{n\ell X}$, $C_{n\ell X}$ and $b^{\dag}_{{\bf q}\,j}$, 
$b_{{\bf q}\,j}$  stand for the creation and annihilation operators of the 
electron and phonon respectively. The matrix element of electron-phonon 
interaction is given by the formula
\begin{equation}
M^{{\bf q}\,j}_{X}
\!\!\,\left(\!\! 
\begin{array}{c}
n'  \ell' \vspace{-8 pt} \\
n \: \ell
\end{array}
\!\!\right)=(e\beta_{{\bf q}\,j}-iq\Xi_{{\bf q}\,j})\sqrt{\frac{\hbar}
{2\omega_{{\bf q}j}\,\rho_cV_{3d}}}F_{nn'}(q_z)\Omega_{\ell\ell'}(q_{\bot})
e^{iq_x(X-\lambda^2q_y/2)+i\varphi(\ell-\ell')}\;\;,
\label{eq:4}
\end{equation}
where $q_{\bot}=\sqrt{q_x^2+q_y^2}$ is the value of the in-plane component of 
the phonon wave vector {\bf q}, and $\varphi$ is the 
polar angle of ${\bf q_{\bot}}$, counted from the $x$-axis. The sample volume 
and density are noted by $V_{3d}$ and $\rho_c$, respectively. The factor 
$\Omega_{\ell\ell'}(q_{\bot})$ for $\ell\geq\ell'$ reads
\begin{equation}
\Omega_{\ell\ell'}(q_{\bot})=i^{\ell-\ell'}\left(\frac{\ell'!}{\ell!}
\right)^{1/2}e^{-\chi^2/2}\chi^{\ell-\ell'}L^{\ell-\ell'}_{\ell'}(\chi^2)\;\;,
\label{eq:5}
\end{equation}
\begin{displaymath}
\chi^2=\lambda^2q_{\bot}^2/2\;\;,
\end{displaymath} 
where $L_n^m(x)$ are the associated Laguerre polynomials.  For $\ell<\ell'$, 
$\Omega_{\ell\ell'}(q_{\bot})$ is obtained from  
Eq.~(\ref{eq:5}) by interchanging the indices $\ell$ and $\ell'$ in the 
right-hand-side of (\ref{eq:5}). 
The parameter $\Xi_{{\bf q}\,j}$ is introduced as follows,
\begin{equation}
\Xi_{{\bf q}\,j}=\frac{1}{2q}\Xi^{\mu\nu}(q_{\mu}e^{(j)}_{\nu}({\bf q})+
q_{\nu}e^{(j)}_{\mu}({\bf q}))\;\;,
\label{eq:6}
\end{equation} 
$\Xi^{\mu\nu}$ being the deformation potential tensor and 
${\bf e}^{(j)}({\bf q})$ the polarization vector of the $j$-th phonon branch. 
For GaAs, it reduces to $\Xi_{{\bf q}\,j}=\Xi_0\delta_{j1}$, with 
$\Xi_0\sim7\;{\rm eV}$. The parameter $\beta_{{\bf q}\,j}$ is introduced as 
follows,
\begin{equation}
\beta_{{\bf q}\,j}=\frac{4\pi}{2q^2\kappa}\beta^{\mu\nu\omega}q_{\mu}
(q_{\nu}e^{(j)}_{\omega}({\bf q})+q_{\omega}e^{(j)}_{\nu}({\bf q}))\;\;,
\label{eq:7}
\end{equation}
where $\beta^{\mu\nu\omega}$ is the piezomodulus tensor and $\kappa$ is the 
relative permittivity of the material. In GaAs the piezomodulus tensor is 
completely symmetric and has only the off-diagonal components not equal 
to zero. The value of such a component is denoted by 
$h_{14}\sim0.14\;{\rm C/m^2}$ and is called the only piezomodulus of the 
crystal. The parameter $\beta_{{\bf q}\,j}$ thus depends only on the spatial 
orientation of the vector ${\bf q}$ and can be presented in the following 
form
\begin{displaymath}
\beta_{{\bf q}\,1}=\beta_{0}\frac{3}{2}\sin^2{\theta}\cos{\theta}
\sin{2\varphi}\;\;,
\end{displaymath}
\begin{equation}
\beta_{{\bf q}\,2}=\beta_{0}\frac{1}{2}\sin{2\theta}\cos{2\varphi}\;\;,
\label{eq:7a}
\end{equation}
\begin{displaymath}
\beta_{{\bf q}\,3}=\beta_{0}\frac{1}{2}(3\cos^2{\theta}-1)\sin{\theta}
\sin{2\varphi}\;\;,
\end{displaymath}
where $\beta_{0}=(4\pi/\kappa)h_{14}$, and $\theta$ is the angle formed by 
${\bf q}$ and $z$-axis. The quantum well form-factor $F_{nn'}(q_z)$ is defined
 as
\begin{equation}
F_{nn'}(q_z)=\int{\psi_{n'}(z)e^{iq_zz}\psi_n(z)}{dz}\;\;,
\label{eq:8}
\end{equation} 
with $\psi_n(z)$ being the $n$-th size quantization level wave function 
(chosen to be real), which is completely determined by the confinding 
potential $V(z)$. Explicit expressions for $F_{nn'}(z)$ for three cases of 
$V(z)$ are given in Appendix~[1]. As a general feature, the factor 
$F_{nn'}(z)=\delta_{nn'}$ at $q_z\ll1/a$, is of the order of unity at 
$q_z\sim1/a$, and rapidly tends to zero at $q_z\gg1/a$, where $a$ is the 
characteristic size measure entering V(z).

The matrix element (\ref{eq:4}) possesses the following symmetry relation
\begin{equation}
M^{{\bf q}\,j}_{X}
\!\!\,\left(\!\! 
\begin{array}{c}
n'  \ell' \vspace{-8 pt} \\
n \: \ell
\end{array}
\!\!\right)=\left[M^{{\bf -q}\,j}_{X-\lambda^2q_y}
\!\!\,\left(\!\! 
\begin{array}{c}
n \: \ell \vspace{-8 pt} \\
n'  \ell'
\end{array}
\!\!\right)\right]^* \;\;.
\label{eq:8a}
\end{equation} 

We calculate $\sigma_{xx}$ starting from Kubo's formula for the conductivity 
tensor~\cite{ref:Kubo}
\begin{equation}
\sigma_{\mu\nu}(\omega)=\frac{1}{V_{2d}}\int_{0}^{\infty}dt\,e^{-i\omega t}
\int_{0}^{\beta}d\lambda\,\langle J_{\nu}(-i\hbar\lambda)J_{\mu}(t)\rangle\;\;,
\label{eq:9}
\end{equation}
which gives the exact amplitude and phase of induced current in an applied 
electric field of frequency $\omega$. Here ${\bf J}(t)$ is the current 
operator in the Heisenberg representation, $V_{2d}$ is the area of the 2D 
plain, and $\beta=1/T$, with $T$ being the temperature measured in energy 
units. The average denoted by brackets in Eq.~(\ref{eq:9}) is carried out in 
the grand canonical ensemble with the density matrix
\begin{equation}
\rho=\frac{1}{Z}\exp\{-\beta(H-\mu N)\}\;\;,
\label{eq:10}
\end{equation}
where $\mu$ is the electron chemical potential, $N$ is the electron number 
operator, and $Z$ is the partition function of the system of electrons and 
phonons.

In the case of the quantizing magnetic field, the events of electron 
scattering are rare if compared to the frequency of the orbital motion 
(cyclotron frequency). One can say that the electron orbital degree of 
freedom is frozen, and the electron scattering occurs between states 
characterised by the the center of orbit coordinates $X$ and $Y$~\cite{ref:LL}.
The electron conductivity then can be deduced to the migration of the 
electron center of orbit~\cite{ref:KHH,ref:KMH}. Mathematically, it results 
in replacing the current operator ${\bf J}$ in formula (\ref{eq:9}) by 
$-e{\bf\dot{R}}$, where ${\bf R}\equiv(X, Y)$ is the radius-vector of the 
electron center of orbit, and adding an antisymmetric tensor with the 
$xy$-component equal to 
$-en_{2D}/B$ to the conductivity tensor, $n_{2D}$ being the 2D electron 
density. The coordinates $X$ and $Y$ cannot be measured simultaneously. 
Their operators satisfy the commutation relation $[X, Y]=i\lambda^2$. Operator 
${\bf R}$ satisfies the following equation of motion,

\begin{equation}
{\bf\dot{R}}=\frac{i}{\hbar}[H_{int}, {\bf R}]\;\;.
\label{eq:11}
\end{equation}

We shall restrict ourselves to the study of the dissipative conductivity only.
 For this purpose we rewrite the $xx$-component of Eq.~(\ref{eq:9}) in an 
equivalent form,
\begin{equation}
\sigma_{xx}=\lim_{\delta\rightarrow +0}\frac{1}{i\omega+\delta}[\phi_{xx}(0)+
\int_{0}^{\infty}e^{-i\omega t-\delta t}\dot{\phi}_{xx}(t)dt]\;\;,
\label{eq:11a}
\end{equation}
where the response function $\phi_{xx}(t)$ is given by
\begin{equation}
\phi_{xx}(t)=e^2\int_{0}^{\beta}\langle\dot{X}(-i\hbar\lambda)\dot{X}(t)
\rangle d\lambda\;\;,
\label{eq:11b}
\end{equation}
with $\dot{X}(t)=\exp(iHt/\hbar)\dot{X}\exp(-iHt/\hbar)$.
An expression for $\dot{X}$ is straightforward from Eq.~(\ref{eq:11}),
\begin{equation}
\dot{X}=-\frac{i}{\hbar}\sum_{ n\ell n'\ell'\atop {\bf q}\,jX}\lambda^2q_y
M^{{\bf q}\,j}_{X}
\!\!\,\left(\!\! 
\begin{array}{c}
n'  \ell' \vspace{-8 pt} \\
n \: \ell
\end{array}
\!\!\right)\, 
(b^{\dag}_{-{\bf q}\,j}+b_{{\bf q}\,j})C^{\dag}_{n'\ell'X-
\lambda^2q_y}C_{n\ell X}\;\;.
\label{eq:12}
\end{equation}
Introducing the retarded bosonic Green's function
\begin{equation}
\Pi^R(t-t')=-i{\mit\vartheta}(t-t')\langle\,[\dot{X}(t)\dot{X}(t')-
\dot{X}(t')\dot{X}(t)]\,\rangle\;\;,
\label{eq:13}
\end{equation}
and noting that
\begin{equation}
\dot{\phi}_{xx}(t)=\frac{ie^2}{\hbar}\langle\,[\dot{X}\dot{X}(t)-
\dot{X}(t)\dot{X}]\,\rangle\;\;,
\label{eq:13a}
\end{equation} 
one can express the dissipative conductivity in the following 
way~\cite{ref:Asihara}
\begin{equation}
\sigma_{xx}=\frac{ie^2}{\hbar V_{2d}}\left.
\frac{\partial\tilde{\Pi}^R(\omega)}{\partial\omega}
\right|_{\omega\rightarrow 0}\;\;,
\label{eq:14}      
\end{equation}
where the Fourier transform of $\Pi^R(t)$ is given by
\begin{equation}
\tilde{\Pi}^R(\omega)=\int_{0}^{\infty}e^{i\omega t}\Pi^R(t)dt\;\;.
\label{eq:15}
\end{equation}
Formula (\ref{eq:14}) deduces the calculation of the dissipative conductivity 
to the evaluation of the two-particle Green's function~(\ref{eq:13}). The 
latter can be easily performed using the finite-temperature diagramatic 
technique~\cite{ref:Mah}, which is based on the Matsubara Green function 
introduced as,
\begin{equation}
\Pi(\tau)=-{\mit\vartheta}(\tau)\langle\, \dot{X}(\tau)\dot{X} 
\,\rangle-{\mit\vartheta}(-\tau)\langle\, \dot{X}\dot{X}(\tau) \,\rangle\;\;,
\label{eq:16}
\end{equation}
where $\tau$ is the imaginary time ($-\beta\le\tau\le\beta$), and 
$\dot{X}(\tau)$, in this formula, is defined as  
\begin{equation}
\dot{X}(\tau)=e^{\tau(H-\mu N)}\dot{X}e^{-\tau(H-\mu N)}\;\;.
\label{eq:17}
\end{equation}
Performing an S-matrix expansion of the Green's function $\Pi(\tau)$ and 
expressing each term in terms of non-perturbed electron and phonon Green's 
functions, one obtaines an infinite series of diagrams, the first terms of 
which are shown on Fig.~2. This series can be summed up graphically to the 
diagram of Fig.~{3{\it a}}. The solid heavy lines mean the exact electron 
Green's function $G_{X}(n\ell;\,ip_n)$, and the dashed heavy line represents 
the exact phonon Green's function $D({\bf q}j;\, i\omega_m)$. The shaded 
triangle is the total vertex part of the electron-phonon interaction, which 
we note by 
${\cal M}_{X}^{{\bf q}j}\Big({n'\ell\,' \atop n\ell}; ip_n, i\omega_n\Big)$.

The Fourier transform $\tilde{\Pi}^R(\omega)$ is obtained from the Fourier 
transform of $\Pi(\tau)$ by the standart analytical continuation to the real 
axis~\cite{ref:AGD}, {\it i. e.} by the replacement 
$i\omega_n\longrightarrow\omega+i\delta$.
An analytic expression for the Fourier transform of $\Pi(\tau)$ follows form 
the diagram of Figure~{3{\it a}},
\begin{eqnarray}
\tilde{\Pi}(i\omega_n)&=&-\frac{\lambda^4}{\hbar^2\beta^2}\!\!
\sum_{ n\ell n'\ell'\atop {\bf q}\,jX}\sum_{i\omega_m}\sum_{ip_n}
M^{{\bf q}\,j}_{X}
\!\!\,\left(\!\! 
\begin{array}{c}
n'  \ell' \vspace{-8 pt} \\
n \: \ell
\end{array}
\!\!\right)\,\hat{Q}_y^2\,{\cal M}^{-{\bf q}\,j}_{X-\lambda^2q_y}
\!\!\,\left(\!\! 
\begin{array}{c}
n\:  \ell \vspace{-8 pt} \\
n' \ell'
\end{array}; ip_n, i\omega_m
\right)\times\nonumber\\
&&D({\bf q}j;\, i\omega_m)G_{X-\lambda^2q_y}(n'\ell\,';\,ip_n+
i\omega_m-i\omega_n)G_{X}(n\ell;\,ip_n)\;\;,
\label{eq:18}
\end{eqnarray} 
where the operator $\hat{Q}_y$  acts on the total vertex part 
${\cal M}_{X}^{{\bf q}j}\Big({n'\ell\,' \atop n\ell}; ip_n, i\omega_n\Big)$ 
and chooses the component $q_y$ referring to the non-perturbed phonon line of 
one of the fermionic angles of the total vertex. The Matsubara frequency 
$i\omega_n$ has to be inserted in the same vertex where $\hat{Q}_y$ acts.

In the next Section we shall work in a representation where the electron 
Green's function is a matrix. It can be obtained by projecting 
$G_{X}(n\ell;\,ip_n)$ onto the single-particle states of the electron. 

\section{Energy Spectrum of Electrons in the Presence of Phonons}
It is well known that, in the general case, taking into account of a 
perturbation has two effects on a electron gass. One is the renormalisation 
of the energy spectrum, and the other is the introduction of a finite 
lifetime for electrons. We will focus here on the energy renormalisation 
effects in a special case when two Landau levels of different subbands of 
size quantisation have approached each other at an energy distance comparable 
to the characteristic phonon energy. Let 1 (2) be a combined label $n\ell$ 
for the energy level belonging to the lower (upper) size quantisation 
subband, and hence ${\cal E}_1$ and ${\cal E}_2$ stand for the energies of 
these levels.

The renormalised electron energy spectrum can be extracted from the poles of 
the electron Green's function. By the virtue of circumstances the electron 
self-energy is diagonal in the Landau level number, and hence, so is the the 
exact electron Green's function. Therefore, one can consider the Green's 
function of each electron level separately. For the level $1$ in the presence 
of the level $2$, the Dyson equaton reads,
\begin{equation}
G_{11}(ip_n)=G_{11}^{(0)}(ip_n)+G_{11}^{(0)}(ip_n)\Sigma_{11}(ip_n)
G_{11}(ip_n)\;\;,
\label{eq:26}
\end{equation}
where $G_{11}^{(0)}(ip_n)$ is the Green's function of the non-perturbed 
electron of level $1$, $\Sigma_{11}(ip_n)$ is the self-energy of the electron 
level $1$, and $ip_n$ is the imaginary fermionic frequency, 
$p_n=(2n+1)\pi/\beta$. 
The influence of level $2$ is manifested indirectly through the self energy 
$\Sigma_{11}(ip_n)$. The lowest order approximation for $\Sigma_{11}(ip_n)$ 
is given by the diagram of Fig.~4. An analytical expression of this diagram 
is as follows,
\begin{equation}
\Sigma_{11}^{(0)}(ip_n)=\Sigma_{11}^{\mbox{\scriptsize\it self}}(ip_n)+
\Sigma_{11}^{\mbox{\scriptsize\it inter}}(ip_n)\;\;,
\label{eq:16a}
\end{equation}
with
\begin{equation}
\Sigma_{11}^{\mbox{\scriptsize\it self}}(ip_n)=\sum_{{\bf q}\,j}
\left|M^{{\bf q}\,j}_{X}
\!\!\,\left(\! 
\begin{array}{c}
1 \vspace{-6 pt} \\
1
\end{array}
\!\right)\right|^2\left(\frac{N_{\bf q}+
n_F(\xi_{1})}{ip_n+\omega_{{\bf q}\,j}-\xi_{1}}+\frac{N_{\bf q}+1-
n_F(\xi_{1})}{ip_n-\omega_{{\bf q}\,j}-\xi_{1}}\right)
\label{eq:27}\;\;,
\end{equation}
and
\begin{equation}
\Sigma_{11}^{\mbox{\scriptsize\it inter}}(ip_n)=\sum_{{\bf q}\,j}
\left|M^{{\bf q}\,j}_{X}
\!\!\,\left(\! 
\begin{array}{c}
2 \vspace{-6 pt} \\
1
\end{array}
\!\right)\right|^2\left(\frac{N_{\bf q}+
n_F(\xi_{2})}{ip_n+\omega_{{\bf q}\,j}-\xi_{2}}+\frac{N_{\bf q}+1-
n_F(\xi_{2})}{ip_n-\omega_{{\bf q}\,j}-\xi_{2}}\right)\;\;,
\label{eq:28}
\end{equation}
where $\xi_i={\cal E}_i-\mu$. The self-energy part 
$\Sigma_{11}^{\mbox{\scriptsize\it self}}$ accounts for the renormalization 
of the energy level in the absence of other levels. This 
renormalization is not expected to change significantly in the neighbourhood 
of the level crossing, and we will not take it into account, considering that 
it has been included initially in the electron effective mass. This assumption
 is correct as far as the deviation from the non-perturbated energy levels is 
small in comparison with the characteristic phonon energy.

Neglecting the imaginary part of the retarded self-energy, which is equivalent
 to using the Brillouin-Wigner perturbation theory~\cite{ref:Mah}, one 
arrives at the following equation for the renormalized electron energy level,
\begin{equation}
E-{\cal E}_1-\Sigma_{11}^{\mbox{\scriptsize\it inter}}(E)=0\;\;.
\label{eq:29}
\end{equation}
Solving Eq.~(\ref{eq:29}) relative to $E$ one finds the renormalized electron 
energy level $\tilde{\cal E}_1$. A similar equation holds for the level 2.

Fig.~5 shows the qualitative dependence of the renormalized energy levels 
$\tilde{\cal E}_1$ and $\tilde{\cal E}_2$ on the magnetic field $B$ for the 
crossing of the Landau levels $\ell=1$ and $\ell=0$ of the first and the 
second subbands of size quantization, respectively. One can see that there 
are two values of $B$ on both sides of the crossing, where repulsion of 
levels changes to attraction. We can prove analytically that this happens 
when the energy distance betwen the two unperturbed electron levels is close 
to the caracteristic phonon energy $\hbar s/\lambda_+$, where $\lambda_+$ is 
$\lambda$ at the level crossing. This characteristic energy of phonons will 
reappear in Section~IV in the calculation of the dissipative conductivity 
$\sigma_{xx}$, where some peculiar behaviour of $\sigma_{xx}$ are found at 
the same characteristic magnetic fields on both sides of the level crossing. 

Notably, the typical value of the deviation $\delta E$ of the renormalized 
levels from unperturbed is very small for all temperatures of interest in 
GaAs systems. A rough analytical estimate gives $\delta E\sim\alpha T$ at 
temperatures $T\gg\hbar s/\lambda_+$, but still low enough that the 
perturbation theory works, $T\ll\hbar s/(\alpha\lambda_+)$. The dimensionless 
coupling constant of electron-phonon interaction 
$\alpha\simeq0.1[(100\:{\rm \AA})/\lambda]^2$ for GaAs~\cite{ref:GL}. 
Thus, for 
$B=6.6$~Tesla and $T=10\;K$ one obtains $\delta E\sim10^{-1}\;{\rm\mbox meV}$.
 Numerical calculations give an even smaller value of 
$\delta E\sim10^{-3}\;{\rm\mbox meV}$. We shall neglect these energy 
corrections in our further calculations, stating only that the electron 
levels do intersect in the presence of equilibrium phonons.

\section{DISSIPATIVE CONDUCTIVITY}
The dissipative conductivity $\sigma_{xx}$ is calculated using formula 
(\ref{eq:14}) and the expression for the Fourier transform of the 
two-particle correlation function (\ref{eq:18}). 
The results of the perturbation theory are obtained using the 
S-matrix expansion for the electron Green's functions. 
The effects of renormalization of the phonon spectrum due 
to the presence of electrons will not be taken into account in what follows. 
We have made a rougher approximation already, neglecting the interface phonon 
modes and the variance of the sound velocity, as one goes from the quantum 
well to the substrate material. We use the Green's function of non-perturbed 
phonons instead of the exact phonon Green's function.

\subsection{The One-phonon Process}
This process is described by the first diagram of Fig.~2. The analytical 
expression of this diagram reads,
\begin{displaymath}
\tilde{\Pi}_0(i\omega_n)=-\frac{e^2}{\hbar^2}\lambda^4\frac{1}{\beta}
\sum_{i\omega_m}\frac{1}{\beta}\sum_{ip_n}
\sum_{ n\ell n'\ell'\atop {\bf q}\,jX}
q_y^2\left|M^{{\bf q}\,j}_{X}
\!\!\,\left(\!\! 
\begin{array}{c}
n'  \ell' \vspace{-6 pt} \\
n \: \ell
\end{array}
\!\!\right)\right|^2\times
\end{displaymath}
\vspace{-24pt}
\begin{equation}
D^{(0)}({\bf q}j;\, i\omega_m)G_{X-\lambda^2q_y}^{(0)}
(n'\ell\,';\,ip_n+i\omega_m-i\omega_n)G_{X}^{(0)}(n\ell;\,ip_n)\;\;,
\label{eq:30}
\end{equation}
where $G_X^{(0)}(n\ell, ip_n)=1/(ip_n-\xi_{n\ell})$ is the non-perturbed 
electron Green's function, and 
$D^{(0)}({\bf q}j, i\omega_m)=-2\hbar\omega_{{\bf q}j}/(\omega_m^2+\hbar^2\omega_{{\bf q}j}^2)$ 
is the non-perturbed phonon Green's function. The summation over the 
discrete Matsubara frequencies can be easily performed in (\ref{eq:30}) and, 
using (\ref{eq:14}), one arrives at the following expression for the 
dissipative conductivity,
\begin{displaymath}
\sigma_{xx}=\frac{1}{V_{2d}}\frac{e^2}{\hbar^2}\lambda^4\pi\beta
\sum_{ n\ell n'\ell'\atop {\bf q}\,jX}
q_y^2\left|M^{{\bf q}\,j}_{X}
\!\!\,\left(\!\! 
\begin{array}{c}
n'  \ell' \vspace{-8 pt} \\
n \: \ell
\end{array}
\!\!\right)\right|^2f(\xi_{n\ell})\times
\end{displaymath}
\vspace{-18pt}
\begin{displaymath}
\left\{N_{{\bf q}j}[1-f(\xi_{n\ell}+\hbar
\omega_{{\bf q}j})]\delta(\xi_{n'\ell\,'}-\xi_{n\ell}-
\hbar\omega_{{\bf q}j})+\right.
\end{displaymath}
\vspace{-24pt}
\begin{equation}
\left.(N_{{\bf q}j}+1)[1-f(\xi_{n\ell}-\hbar\omega_{{\bf q}j})]
\delta(\xi_{n'\ell\,'}-\xi_{n\ell}+\hbar\omega_{{\bf q}j})\right\}\,\,,
\label{eq:31}
\end{equation}
where $f(\xi)=1/(\exp(\beta\xi)+1)$ is the fermionic distribution function 
and $N_{{\bf q}j}=1/(\exp(\beta\hbar\omega_{{\bf q}j})-1)$ is the bosonic 
distribution function. 

Formula (\ref{eq:31}) has the following physical interpretation. The electron 
can be scattered from the lower (upper) in energy Landau level onto the other 
one by absorbing (emitting) a phonon. Each of these processes leads to the 
diffusion of the electron within these two Landau levels. Thus, one can think 
of the diffusion of the electron in the upper level induced by the emission of 
a phonon, and of the diffusion of the electron in the lower level induced by 
the absorption of a phonon. At thermal equlibrium the number of electrons 
transfered up and down is equal, which mathematically is related by the 
identity,
\begin{equation}
f({\cal E}_1)(1-f({\cal E}_2))N_{q_0}=f({\cal E}_2)(1-f({\cal E}_1))
(N_{q_0}+1)\;,
\label{eq:32}
\end{equation}
where $q_0=|{\cal E}_2-{\cal E}_1|/(\hbar s)$. Threfore, it is not important 
that at each scattering event the electron is transfered from one 
Landau level onto the other. In a diffusive motion the ``memory'' after one 
scattering event is lost. Thus, one can safely assign to each Landau level 
a diffusion coefficient and disregard the conservation of the number of 
electrons during inelastic processes. This way we interpret formula 
(\ref{eq:31}) as a generalised Einschtein relation,
\begin{equation}
\sigma_{xx}=\frac{e^2}{2\pi\lambda^2}\frac{f({\cal E}_1)}{T}D_{xx}^{1\to 2}+
\frac{e^2}{2\pi\lambda^2}\frac{f({\cal E}_2)}{T}D_{xx}^{2\to 1}\;,
\label{eq:31a}
\end{equation}
with the diffusion coefficients being calculated according to the formula,
\begin{equation}
D_{xx}^{\alpha\to\beta}=\frac{1}{2}\sum_{{\bf q}j}\sum_{X'}(X-X')^2W_{XX'}^{\alpha\to\beta}\;.
\label{eq:33}
\end{equation}
The probability $W_{XX'}^{\alpha\to\beta}$ for an electron in level $\alpha$ 
to be scattered from point $X$ to point $X'$, and at the same time to be 
transfered to the level $\beta$, can be as well calculated with the means of 
the Fermi golden-rule, which straightforwardly yields,
\begin{eqnarray}
&&W_{XX'}^{1\to 2}=\frac{2\pi}{\hbar}\left|M_{X}^{{\bf q}j}\left({2\atop 1}
\right)\right|^2(1-f({\cal E}_2))N_{{\bf q}j}\delta_{X',X-\lambda^2q_y}
\delta({\cal E}_2-{\cal E}_1-\hbar\omega_{{\bf q}j})\;,\nonumber\\
&&W_{XX'}^{2\to 1}=\frac{2\pi}{\hbar}\left|M_{X}^{-{\bf q}j}\left({1\atop 2}
\right)\right|^2(1-f({\cal E}_1))(N_{-{\bf q}j}+1)\delta_{X',X-\lambda^2q_y}
\delta({\cal E}_2-{\cal E}_1+\hbar\omega_{-{\bf q}j})\;.
\label{eq:34}
\end{eqnarray}
The processes of phonon absorption and emission give equal contributions into 
$\sigma_{xx}$. Formula (\ref{eq:31a}) together with (\ref{eq:32}),(\ref{eq:33})
give just the same result for $\sigma_{xx}$ as formula (\ref{eq:31}) when the 
contribution from the levels other than the two considered ones is neglected 
in (\ref{eq:31}). Finally, we note that such a simple picture of independent 
diffusion of electrons in each Landau level is not valid generally in a 
non-equilibrium situation.

Taking into account the contributions from the deformation potential ($dp$), 
longitudinal-phonon piezo-electric (peLA), and transverse-phonon 
piezo-electric (peTA) mechanisms of electron-phonon interaction, one obtains,
\begin{equation}
\sigma_{xx}=\sigma_{xx}^{dp}+\sigma_{xx}^{peLA}+\sigma_{xx}^{peTA}\;\;,
\label{eq:50}
\end{equation}
where
\begin{eqnarray}
\sigma_{xx}^{dp}&=&\frac{\sqrt{2}}{\pi^2}\frac{e^2}{\hbar}
\frac{\Xi_{q_{_{0LA}}}^2\hbar}{2\rho_cs_{_{LA}}\lambda^4}
\frac{1}{T_{\lambda}T}f(\xi_1)(1-f(\xi_2))N_{q_{_{0LA}}}
\left(\frac{q_{_{0LA}}\lambda}{\sqrt{2}}\right)^{5+2(\ell-\ell')}\nonumber\\
&\times&\frac{\ell'!}{\ell!}\int_{0}^{1}dxZ(q_{_{0LA}}x)
(1-x^2)^{\ell-\ell'+1}\left[L_{\ell'}^{\ell-\ell'}
((1-x^2)q_{_{0LA}}^2\lambda^2/2)\right]^2e^{-(1-x^2)
q_{_{0LA}}^2\lambda^2/2}\;\;,
\end{eqnarray}
and
\begin{displaymath}
\sigma_{xx}^{peLA}=\frac{9}{8\sqrt{2}\pi^2}\frac{e^2}{\hbar}
\frac{e^2\hbar\beta_{q_{_{0LA}}}^2}{2\rho_cs_{_{LA}}\lambda^2}
\frac{1}{T_{\lambda}T}f(\xi_1)(1-f(\xi_2))N_{q_{_{0LA}}}
\left(\frac{q_{_{0LA}}\lambda}{\sqrt{2}}\right)^{3+2(\ell-\ell')}\times
\end{displaymath}
\vspace{-12pt}
\begin{equation}
\frac{\ell'!}{\ell!}\int_{0}^{1}dxZ(q_{_{0LA}}x)x^2(1-x^2)^{\ell-\ell'+3}
\left[L_{\ell'}^{\ell-\ell'}((1-x^2)q_{_{0LA}}^2\lambda^2/2)\right]^2
e^{-(1-x^2)q_{_{0LA}}^2\lambda^2/2}\;\;,
\end{equation}
and
\begin{displaymath}
\sigma_{xx}^{peTA}=\frac{1}{8\sqrt{2}\pi^2}\frac{e^2}{\hbar}
\frac{e^2\hbar\beta_{q_{_{0TA}}}^2}{2\rho_cs_{_{TA}}\lambda^2}
\frac{1}{T_{\lambda}T}f(\xi_1)(1-f(\xi_2))N_{q_{_{0TA}}}\left
(\frac{q_{_{0TA}}\lambda}{\sqrt{2}}\right)^{3+2(\ell-\ell')}\times
\end{displaymath}
\vspace{-12pt}
\begin{equation}
\frac{\ell'!}{\ell!}\int_{0}^{1}dxZ(q_{_{0TA}}x)(9x^4-2x^2+1)
(1-x^2)^{\ell-\ell'+2}\left[L_{\ell'}^{\ell-\ell'}((1-x^2)q_{_{0TA}}^2
\lambda^2/2)\right]^2e^{-(1-x^2)q_{_{0TA}}^2\lambda^2/2}\;\;.
\end{equation}
Here $q_{_{0j}}=({\cal E}_2-{\cal E}_1)/\hbar s_j$, and the form-factor 
$Z(q_z)=|F_{n_1n_2}(q_z)|^2$ depends on the form of the quantum well 
potential $V(z)$.

The largest value of $\sigma_{xx}$ is obtained for the crossing of the Landau 
levels with the smallest possible numbers, {\it i.e.} $\ell=1$ and $\ell'=0$, 
and belonging to the two lowest subbands of size quantization. For the case of 
a parabolic quantum well of characteristic size $a$ ($E_{12}=\hbar^2/ma^2$) 
the calculations can be performed analytically and the result reads,
\begin{equation}
\sigma_{xx}^{dp}=\frac{e^2}{h}A\alpha_{dp}\frac{T_{\lambda}}{T}f(\xi_1)
(1-f(\xi_2))N_{q_{_{0LA}}}\lambda^7a^2q_{_{0LA}}^9
\exp\left(-\frac{\lambda^2q_{_{0LA}}^2}{2}\right)\;\;,
\label{eq:61}
\end{equation}

\begin{equation}
\sigma_{xx}^{peLA}=\frac{e^2}{h}B\alpha_{peLA}\frac{T_{\lambda}}{T}f(\xi_1)
(1-f(\xi_2))N_{q_{_{0LA}}}\lambda^5a^2q_{_{0LA}}^7
\exp\left(-\frac{\lambda^2q_{_{0LA}}^2}{2}\right)\;\;,
\label{eq:62}
\end{equation}

\begin{equation}
\sigma_{xx}^{peTA}=\frac{e^2}{h}C\alpha_{peTA}\frac{T_{\lambda}}{T}f(\xi_1)
(1-f(\xi_2))N_{q_{_{0TA}}}\lambda^5a^2q_{_{0TA}}^7
\exp\left(-\frac{\lambda^2q_{_{0TA}}^2}{2}\right)\;\;,
\label{eq:63}
\end{equation}
where $A\sim10^{-3}$ is a weak function of the magnetic field. The effective 
coupling constants of electron-phonon ineteraction for differents mechanisms 
were introduced in the following way, 
\begin{equation}
\alpha\equiv\alpha_{dp}=\frac{\hbar\Xi_{_0}^2}{2\rho_cs_{_{LA}}\lambda^4}\frac{1}{T_{\lambda}^2}
\end{equation}
\begin{equation}
\alpha_{peLA}=\frac{e^2\beta_{_0}^2}{2\rho_cs_{_{LA}}^3\hbar}
\end{equation}
\begin{equation}
\alpha_{peTA}=\frac{e^2\beta_{_0}^2}{2\rho_cs_{_{TA}}^3\hbar}
\end{equation}

Fig.~\ref{sigma_xx} shows the dependence of $\sigma_{xx}$ 
(formula (\ref{eq:50})) on 
the magnetic field $B$ in the vicinity of an intersection of two Landau 
levels in a parabolic quantum well. The magnetic field corresponding to the 
intersection of the levels is that of the minimum on the graph. The two 
maxima appear at magetic fields close to those at which the electron-phonon 
scattering rate is maximal. The distance between the Landau levels at the 
maxima of $\sigma_{xx}$ is of the order of $T_{\lambda}=\hbar s/\lambda$. In 
computing the dependence of $\sigma_{xx}$ on the magnetic field, the electron 
concentration $n_{2D}$ was kept constant and the chemical potential $\mu$ was 
calculated according to the following formula,
\begin{equation}
\mu=\frac{{\cal E}_1+{\cal E}_2}{2}-T\ln\left(\frac{\sqrt{1+\zeta^2\sinh^2
\frac{{\cal E}_2-{\cal E}_1}{2T}}-\zeta
\cosh\frac{{\cal E}_2-{\cal E}_1}{2T}}{1+\zeta}\right)
\end{equation}
with
\begin{equation}
\zeta=\frac{\lambda^2}{\lambda_+^2}\nu_0-(N_0+1)
\end{equation}
where $\nu_0$ is the filling factor at $\lambda=\lambda_+$, and $N_0$ is the 
number of Landau levels bellow the two considered ones. The electron 
concentration $n_{2D}$ is related to the filling factor $\nu_0$ in the usual 
way, $n_{2D}=\nu_0/2\pi\lambda_+^2$.

We compare different contributions into $\sigma_{xx}$ (see Eq.~(\ref{eq:50}))
on Fig.~\ref{sigma_dppe}. At a small value of the quantum well width $a$,
the deformation potential mechanism (dashed line) dominates the dissipative 
conductivity (Fig.~\ref{sigma_dppe}$a$). However, at a larger value of $a$,
the piezoelectric mechanism is the dominant contribution into $\sigma_{xx}$,
e.g. the piezoelectric interaction with the transversal phonon mode
(solid line) in Fig.~\ref{sigma_dppe}$b$. This is not surprising, since
the deformation potential mechanism contains one extra power of the 
phonon momentum, as it is seen from the matrix element (\ref{eq:4}).
The relevant phonon momentum is related to the quantum well width and to the
magnetic length, $q\sim a\sim \lambda$. The characteristic well width,
at which the crossover from the deformation potential to the piezoelectric
mechanism happens, is given by $a\sim \Xi_0\kappa/eh_{14}$. We find that, 
although this criterion gives a characteristic $a\sim 700\:{\rm\AA}$, the 
actual crossover happens at a smaller value of $a$ 
(see Fig.~\ref{sigma_dppe}$b$), owing to a larger overlap integral of the 
electron with the phonons via the piezoelectric interaction.

\subsection{The Two-phonon Process}
When the distance between the two Landau levels on both sides of the Fermi 
level is much greater than $\hbar s/\lambda$, then the one-phonon electron 
transitions become inefficient, and the two-phonon ones should be taken into 
account. The two-phonon process associated with the electron transition 
between two Landau levels was considered in Ref.~\onlinecite{ref:Falko}. The 
two-phonon process assiciated with the scattering of the electron within the 
same infinitely narrow Landau level was considered in Ref.~\onlinecite{ref:4}.
 Since there was a mistake in Ref.~\onlinecite{ref:4}, we reconsider the 
latter process here again.

Consider one Landau level only, which has a $\delta$-functional density of 
states. Within this infinitely narrow energy level, the one-phonon electron 
transitions are not possible, since phonons with zero energy are not 
efficient for the electron scattering, and, moreover, there are no such 
phonons, because the phonon density of states is proportional to 
$\omega_{{\bf q}j}^2$. However, the multiple-phonon scattering makes it 
possible to transfer momentum without transferring any 
energy to the electron. In the simplest case a simultaneous absorption of one 
phonon with momentum ${\bf q}$ and emission of another phonon with momentun 
${\bf q'}$ transfers a momentum ${\bf q}-{\bf q'}$ to the electron, and 
doesn't change its energy, provided $|{\bf q}|=|{\bf q'}|$.

All the two-phonon processes are described by the diagrams ({\it b})-({\it e}) of Fig.~2. The main contribution is given by the 
processes which do not involve the transition of the electron to another 
Landau level.  The diagram ({\it b}) accounts for the renormalisation of the 
phonon spectrum and will not be taken into account in this paper.
The dissipative conductivity $\sigma_{xx}$ can be presented in the following 
form,
\begin{equation}
\sigma_{xx}=\frac{e^2}{2\pi\lambda^2}f(\xi)(1-f(\xi))\frac{D}{T}
\label{eq:71}
\end{equation}
The diffusion coefficient for the deformation potential type of interaction:
\begin{displaymath}
D^{dp}=\frac{\Xi_0^4\lambda^4}{2\rho_c^2s_{_{LA}}^5\hbar^2}
\frac{1}{(2\pi)^5}\int d^3q\int d^3q'(q_y-q'_y)^2N_q(1+N_{q'})\delta(q-q')
\times
\end{displaymath}
\vspace{-24pt}
\begin{equation}
[L_{\ell}(\lambda^2q_{\bot}^2/2)L_{\ell}(\lambda^2q_{\bot}^{'2}/2)]^2
|Z(q_z)Z(-q'_z)|^2e^{-\lambda^2(q_{\bot}^2+q_{\bot}^{'2})/2}
\sin^2\{\lambda^2{\bf e}_z[{\bf q}_{\bot}\times{\bf q}'_{\bot}]/2\}\;\;,
\label{eq:72}
\end{equation}
where ${\bf e}_z$ is the unit vector normal to the quantum well plane.

At $T\ll T_{\lambda}$ the diffusion coefficient is given by
\begin{equation}
D=A_1\alpha^2\lambda s\left(\frac{T}{T_{\lambda}}\right)^{11}\;\;,
\end{equation}
where $A_1=2^9\pi^7/297\approx5.2\times10^3$.

At $T\gg T_{\lambda}$ the diffusion coefficient is given by
\begin{equation}
D=A_2\alpha^2\lambda s\left(\frac{T}{T_{\lambda}}\right)^2\;\;,
\end{equation}
\begin{equation}
A_2=\frac{3}{2^{9/2}\pi^{5/2}}\int_{0}^{1}\frac{dx}{(1+x)^{3/2}}
[1-F(5/4,7/4;1;-4x/(1+x)^2)]K(\sqrt{x})\;\;,
\end{equation}
where $K(z)$ is the elliptic integral and $F(\alpha, \beta; n; z)$ is the 
hypergeometric function. The numerical value of the constant $A_2$ is 
approximately $5.7\times10^{-3}$.

%\section{MULTI-PHONON SCATTERING}

\section{CONCLUSIONS}
In this paper we considered the scattering of electrons by equilibrium phonons 
in a two-subband two-dimensional electron gas in a quantizing magnetic field. 
A resonat enhancement of the dissipative conductivity was found in the 
vicinity of an intersection of two Landau levels, belonging to different 
size-quantization subbands.

%Level mixing results in removal of selection rules for cyclotron resonance, 
%{\it e. g.} intersubband transitions which are forbidden for the light 
%polarized in the QW plane become allowed. 

\newpage
\begin{figure}
\narrowtext
{\epsfxsize=10.cm
\centerline{\epsfbox{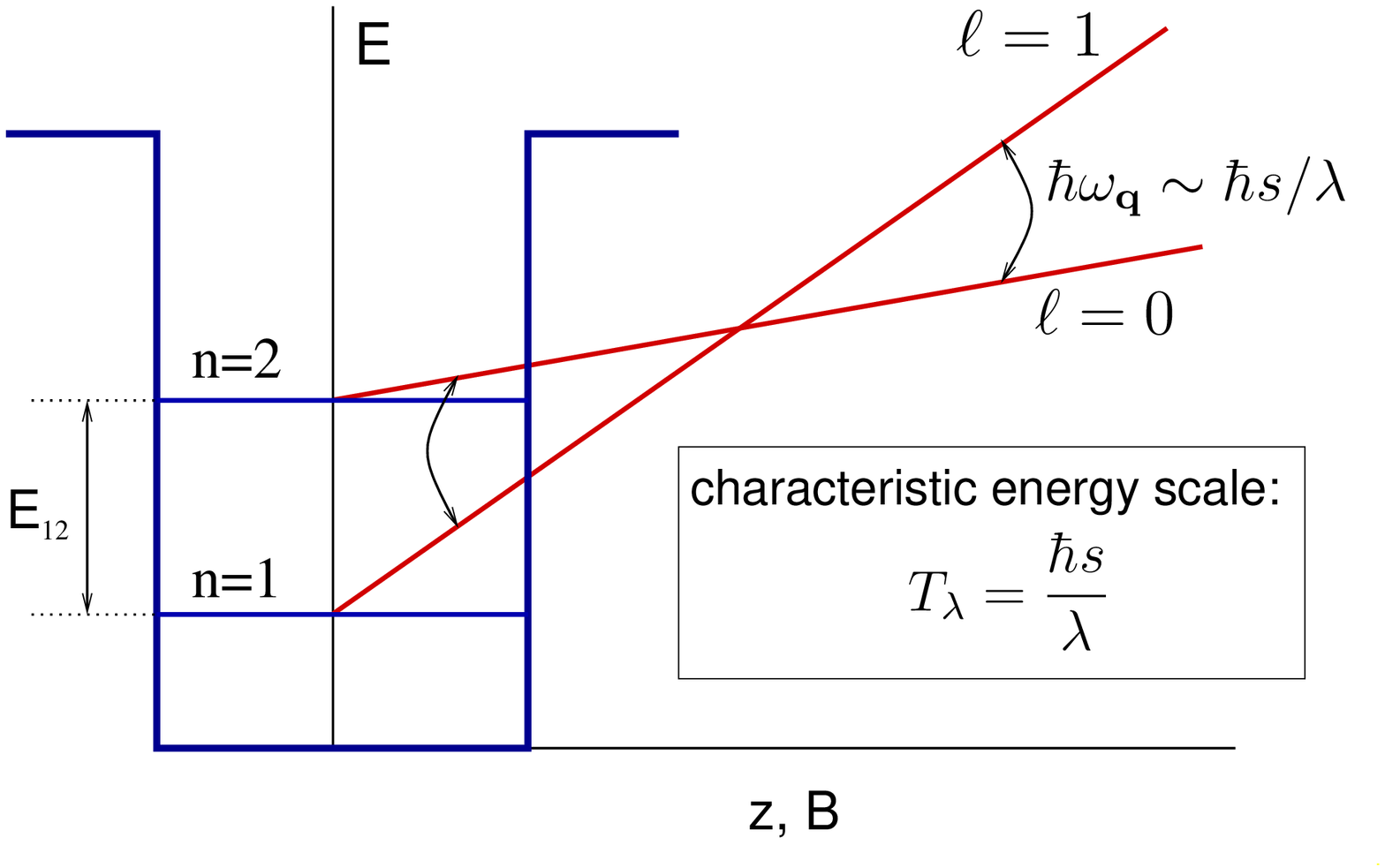}}}
\caption{ A crossing of two Landau levels belonging to different subbands of 
size quantisation. Phonon transitions are most favorable for phonons with 
$q\sim 1/\lambda$.}
\label{fig1}
\end{figure}

\begin{figure}
\narrowtext
{\epsfxsize=15.cm
\centerline{\epsfbox{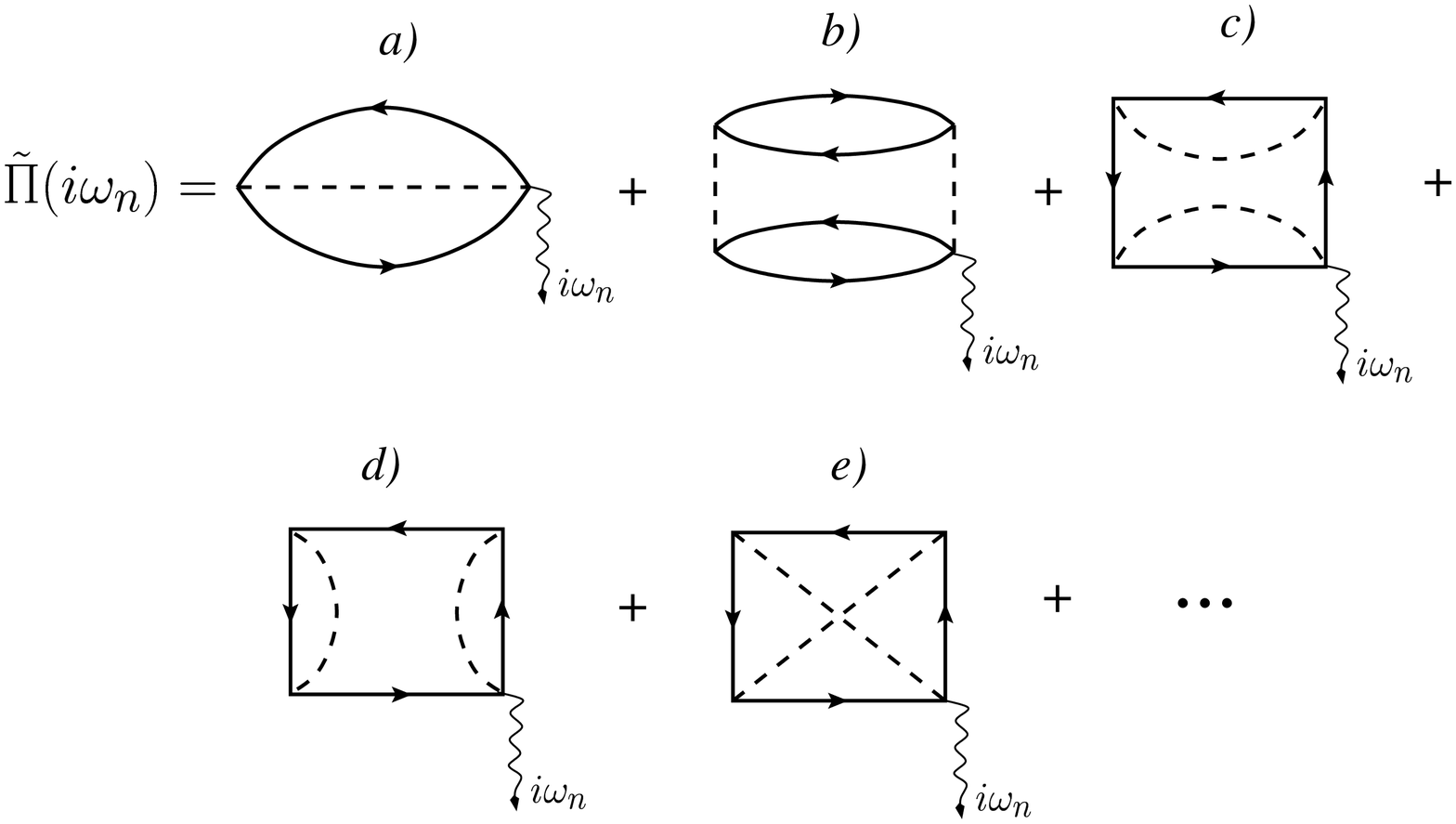}}}
\caption{ S-matrix expantion of the two-particle Green's function $\Pi(\tau)$.
 The solid line and the dashed line represent the Green's functions of the 
non-perturbed electron and phonon respectively. The wavy line was introduced 
to account for the Matsubara frequency $i\omega_n$ and for the action of the 
operator $\hat{Q}_y$.}
\label{fig2}
\end{figure}

\begin{figure}
\narrowtext
{\epsfxsize=8.cm
\centerline{\epsfbox{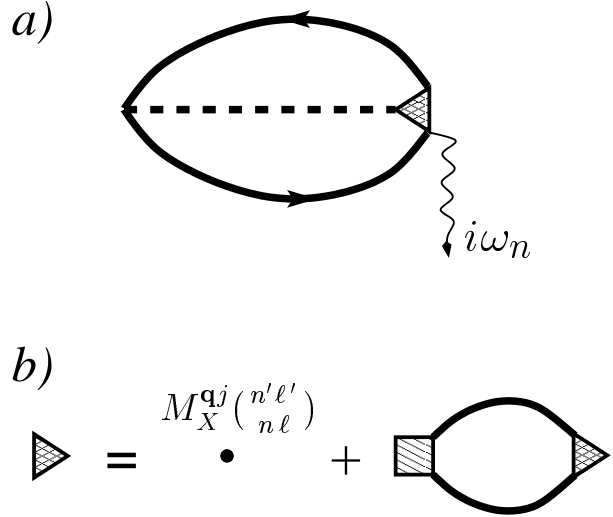}}}
\caption{Results of graphical summation of the perturbation series expansion. 
{\it a}) A diagram for the exact $\tilde{\Pi}(i\omega_n)$ expressed in 
terms of the exact electron Green's function (bold solid line), the exact 
phonon Green's function (bold dashed line), and the total vertex 
part (shaded triangle). {\it b}) A Dyson-type equation for the 
total vertex part, deducing the evaluation of the total vertex part to the 
problem of electron-electron interaction via exchange of phonons. 
The shaded square is the total vertex part for this problem.}
\label{fig3}
\end{figure}

\begin{figure}
\narrowtext
{\epsfxsize=8.cm
\centerline{\epsfbox{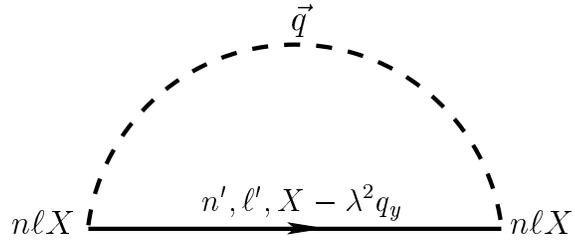}}}
\caption{The first term of the pertubration expansion of the electron self 
energy.}
\label{fig4}
\end{figure}

\begin{figure}
\narrowtext
{\epsfxsize=8.cm
\centerline{\epsfbox{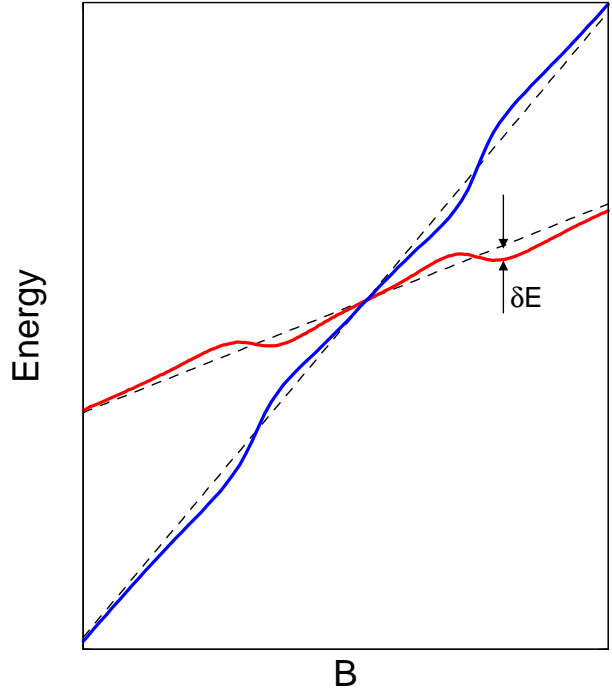}}}
\caption{ Qualitative behaviour of the Landau levels in the vicinity of a 
level-crossing. The dashed lines represent the non-perturbed Landau levels.
}
\label{fig5}
\end{figure}

\begin{figure}
\narrowtext
{\epsfxsize=12.cm
\centerline{\epsfbox{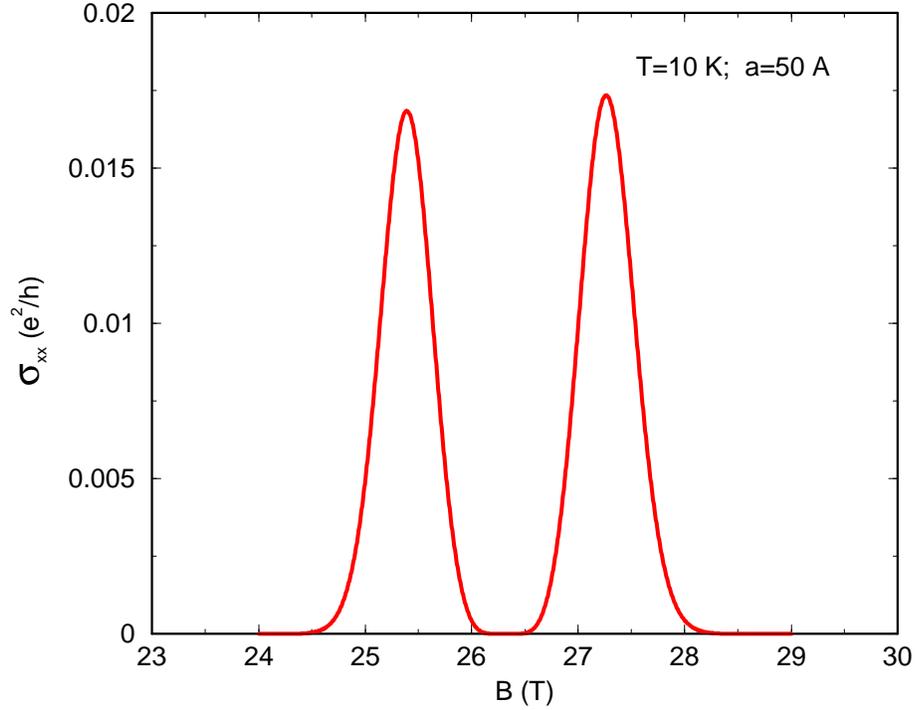}}}
\caption{The dissipative conductivity as a function of magnetic field in 
the vicinity of crossing of the Landau levels $\ell=1$ and $\ell=0$ 
belonging to the first and second size-quantization subbands, respectively. Material parameters are taken as for a GaAs system.}
\label{sigma_xx}
\end{figure}

\begin{figure}
\narrowtext
{\epsfxsize=15.cm
\centerline{\epsfbox{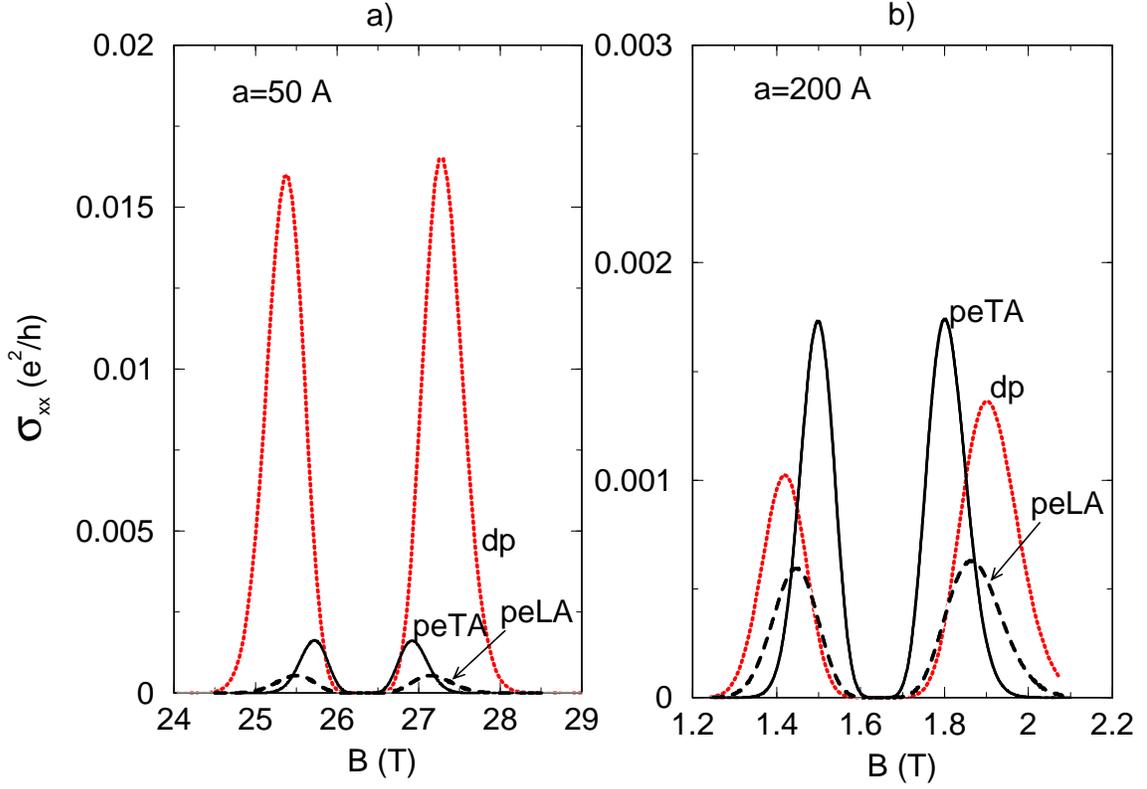}}}
\caption{A comparison of contributions into $\sigma_{xx}$ from different 
mechanisms of electron-phonon interaction at two values of the quantum well 
width $a$.  The lines are labeled as follows: the deformation potential 
mechanism - {\it dp}, the piezoelectric longitudinal mechanism -  {\it peLA}, 
and the piezoelectric transversal mechanism - {\it peTA}.}
\label{sigma_dppe}
\end{figure}

\begin{figure}
\narrowtext
{\epsfxsize=14.cm
\centerline{\epsfbox{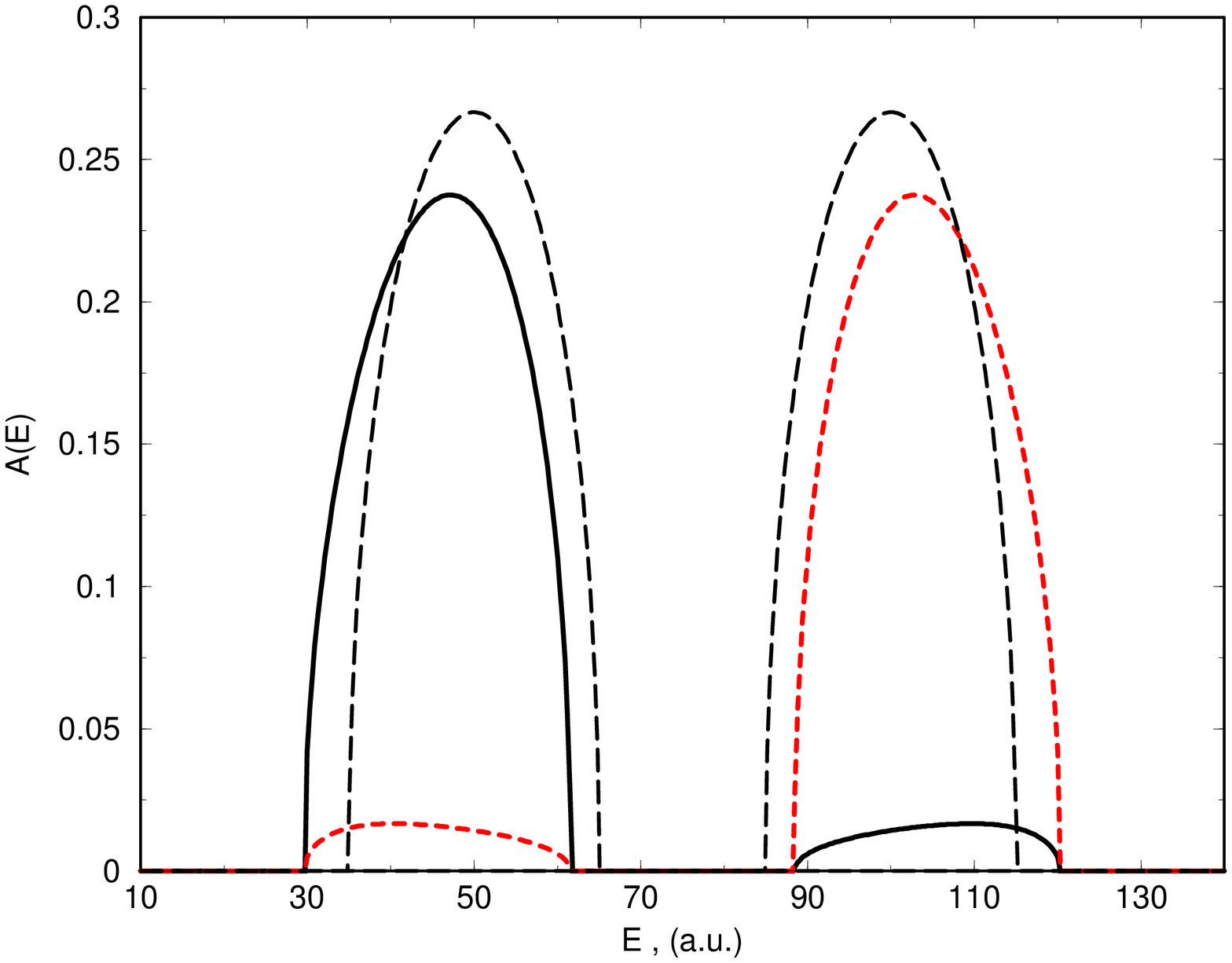}}}
\caption{Level mixing.}
\label{fig8}
\end{figure}


\begin{thebibliography}{99}
\bibitem{ref:1} M.~Sh.~Erukhimov, Fiz. Tekh. Poluprovodn. {\bf 3}, 194 (1969) 
[Sov. Phys. Semicond. {\bf 3}, 162 (1969)].
\bibitem{ref:3} V.~V.~Korneev, Fiz. Tverd. Tela (Leningrad) {\bf 19}, 357 
(1977) [Sov. Phys. Solid State {\bf 19}, 205 (1977)].
\bibitem{ref:KB} K.~A.~Benedict, R.~K.~Hills, and C.~J.~Mellor, Phys. Rev. B 
{\bf 60}, 10~984 (1999).
\bibitem{ref:1a} K.~A.~Benedict, J.~Phys.: Condens. Matter {\bf 3}, 1279 
(1991).
\bibitem{ref:1b} G.~A.~Toombs, F.~W.~Sheard, D.~Neilson, andL.~J.~Challis, 
Solid State Commun. {\bf 64}, 577 (1987). 
\bibitem{ref:Falko} V.~I.~Fal'ko and L.~J.~Challis, J.~Phys.: Condens. 
Matter {\bf 5}, 3945 (1993).
\bibitem{ref:Maksym} P.~A.~Maksym, Springer Series in Solid-State Sciences, 
vol.~{\bf 101}, editor: G.~Landwehr (Springer-Verlag Berlin Heidelberg 1992).


\bibitem{Apalkov_Portnoi} V.~M.~Apalkov and M.~E.~Portnoi, cond-mat/0111378;
cond-mat/0111377.

\bibitem{ref:6} I.~M.~Lifshitz, Zh. Eksp. Teor. Fiz. {\bf 33}, 1569 (1960) 
[Sov. Phys. JETP {\bf 11}, 1130 (1969)].
\bibitem{ref:7} N.~N.~Ablyazov, M.~Yu.~Kuchiev, and M.~E.~Raikh, 
Phys. Rev. B {\bf 44}, 8802 (1991). 
\bibitem{ref:8} D.~V.~Lang, in {\it Deep Centers in Semiconductors}, 
edited by S.~T.~Pantelides (Gordon and Breach, New York, 1986), p.~486.
\bibitem{ref:9} A.~Kastalsky and J.~C.~M.~Hwang, Solid State Commun. 
{\bf 51}, 317 (1984).
\bibitem{ref:10} I.~V.~Kukushkin, K.~von~Klitzing, and K.~Ploog, 
Phys. Rev. B {\bf 40}, 4179 (1989). 
\bibitem{ref:11} A.~J.~Turberfield, S.~R.~Haynes, P.~A.~Wright, R.~A.~Ford, 
R.~G.~Clark, J.~F.~Ryan, J.~J.~Harris, and C.~T.~Foxon, Phys. Rev. 
Lett. {\bf 65}, 635 (1990).
\bibitem{ref:12} A.~S.~Plaut, K.~v.~Klitzing, I.~V.~Kukushkin, and K.~Ploog, 
Proceedings of the 20th International Conference on the Physics of 
Semiconductors, Vol. 2, [Editors: E.~M.~Anastassakis, and 
J.~D.~Joannopoulos], p.~1529 (1990).
\bibitem{ref:13} I.~V.~Kukushkin, V.~B.~Timofeev, K.~von~Klitzing, 
and K.~Ploog, {\it Festk\"{o}rperprobleme (Advances in Solid State Physics)}, 
edited by U.~R\"{o}ssler (Pergamon, Braunschweig, 1988), Vol. 28, p. 21.

\bibitem{ref:4} Yu.~A.~Bychkov, S.~V.~Iordanski, and G.~M.~Eliashberg, 
Pis'ma Zh. Eksp. Teor. Fiz. {\bf 34}, 496 (1981) 
[JETP Lett. {\bf 34}, 473 (1981)].

\bibitem{ref:Wolfe} J.~P.~Wolfe, M.~R.~Hauser, Ann. Physik {\bf 4}, 99 (1995).

\bibitem{ref:Kubo} R.~Kubo, J.~Phys. Soc. Japan {\bf 12}, 570 (1957).
\bibitem{ref:LL} L.~D.~Landau, E.~M.~Lifshitz, {\it Quantum Mechanics} 
(Pergamon Press Ltd. 1977).
\bibitem{ref:KHH} R.~Kubo, H.~Hasegawa, and N.~Hashitsume, J. Phys. Soc. 
Japan {\bf 14} 56 (1959).
\bibitem{ref:KMH} R.~Kubo, S.~Miyake, and N.~Hashitsume, 
{\it Solid State Physics} edited by F.~Seitz and D.~Turnbull, 
Vol. 17, p. 169 (Academic Press, 1965).
\bibitem{ref:Mah} See for example: G.~D.~Mahan, {\it Many-Particle Physics} 
(Plenum Press, New York, 1981).
\bibitem{ref:Asihara} A. Asihara, {\it Statistical Physics} 
(Academic Press 1971); K.~Efetov, {\it Supersymmetry in disorder and chaos} 
(Chambridge University Press 1997). 
\bibitem{ref:AGD} A.~A.~Abrikosov, L.~P.~Gor'kov, and Dzyaloshinskii, 
{\it Methods of Quantum Field Theory in Statistical Physics}, 
(Pergamon Press 1965).

\bibitem{ref:GL} Note that $\alpha\sim B$. For more information see: 
V.~F.~Gantmakher and Y.~B.~Levinson, 
{\it Carrier Scattering in Metals and Semiconductors} 
(North-Holland, Amsterdam, 1987).

\end{thebibliography}
\end{document}